\documentclass{article}

\usepackage{arxiv}

\usepackage[utf8]{inputenc} % allow utf-8 input
\usepackage[T1]{fontenc}    % use 8-bit T1 fonts
\usepackage{hyperref}       % hyperlinks
\usepackage{url}            % simple URL typesetting
\usepackage{booktabs}       % professional-quality tables
\usepackage{amsfonts}       % blackboard math symbols
\usepackage{nicefrac}       % compact symbols for 1/2, etc.
\usepackage{microtype}      % microtypography

\usepackage{authblk}

\usepackage[colorinlistoftodos]{todonotes}

\usepackage{graphicx}

\usepackage{amssymb}

\title{BCN20000: Dermoscopic Lesions in the Wild}

\author[1]{Marc Combalia}
\author[2]{Noel C. F. Codella}
\author[3]{Veronica Rotemberg}
\author[4]{Brian Helba}
\author[5]{Veronica Vilaplana}
\author[3]{Ofer Reiter}
\author[1]{Cristina Carrera}
\author[1]{Alicia Barreiro}
\author[3]{Allan C. Halpern}
\author[1]{Susana Puig}
\author[1]{Josep Malvehy}

\affil[1]{\footnotesize Melanoma Unit, Dermatology Department, Hospital Clínic Barcelona, Universitat de Barcelona, IDIBAPS, Barcelona, Spain}
\affil[2]{\footnotesize  IBM Research AI, T J Watson Research Center, Yorktown Heights, NY, USA}
\affil[3]{\footnotesize Dermatology Service, Department of Medicine, Memorial Sloan Kettering Cancer Center, New York, NY, USA}
\affil[4]{\footnotesize Kitware, Clifton Park, NY, USA}
\affil[5]{\footnotesize Signal Theory and Communications, Universitat Politècnica de Catalunya, Barcelona, Spain}

\begin{document}

\maketitle

\begin{abstract}

This article summarizes the BCN20000 dataset, composed of 19424 dermoscopic images of skin lesions captured from 2010 to 2016 in the facilities of the Hospital Clínic in Barcelona. With this dataset, we aim to study the problem of unconstrained classification of dermoscopic images of skin cancer, including lesions found in hard-to-diagnose locations (nails and mucosa), large lesions which do not fit in the aperture of the dermoscopy device, and hypo-pigmented lesions. The BCN20000 will be provided to the participants of the ISIC Challenge 2019 \cite{ISIC2019}, where they will be asked to train algorithms to classify dermoscopic images of skin cancer automatically.

\end{abstract}

\section{Background and Summary}

Skin cancer is one of the most frequent types of cancer and manifests mainly in areas of the skin most exposed to the sun. 
Since skin cancer occurs on the surface of the skin, its lesions can be evaluated by visual inspection. Dermoscopy is a non invasive method which permits visualizing more profound levels of the skin as its surface reflection is removed. Prior research has found that this technique permits improved visualization of the lesion structures, enhancing the accuracy of dermatologists \cite{argenziano2003dermoscopy, kittler2002diagnostic}. 
 
The increased availability of dermoscopic images has motivated the appearance of more sophisticated algorithms based on deep learning, mainly on convolutional neural networks \cite{esteva2017dermatologist, xie2016melanoma, bi2017automatic}. A significant player in the adoption of these algorithms in the community has been the International Skin Imaging Collaboration (ISIC), which has been organizing yearly challenges since 2016, where participants are asked to develop computer vision algorithms to segment and classify skin lesions in dermoscopic images \cite{marchetti2018results, gutman2016skin, codella2018skin, codella2019skin}. Tschandl et al. showed that the performance of expert dermatologist was already surpassed by the top-scoring algorithms of the ISIC 2018 Challenge \cite{tschandl2019comparison, codella2018skin}. However, as the authors already pointed out, the algorithms tended to perform worse on images from other dermoscopic data sources, which were not represented in the HAM10000 dataset \cite{tschandl2018ham10000}.

In BCN20000, we aim to study the problem of unconstrained classification of dermoscopic images of skin cancer, including lesions found in hard to diagnose locations (nails and mucosa), not segmentable and hypopigmented lesions: dermoscopic lesions in the wild.  Most of the images would be considered hard-to-diagnose and had to be excised and histopathologically diagnosed. Together with the images, we provide valuable information related to the anatomic location of the lesion and the age and sex of the patients. Our efforts aim at creating a challenge which is more similar to what the dermatologists are doing when visiting a patient in the clinical practice.

\begin{figure}[!htb]
\centering
\minipage{0.9\textwidth}
    \minipage{1\textwidth}
      \minipage{0.24\textwidth}
        \includegraphics[width=\linewidth]{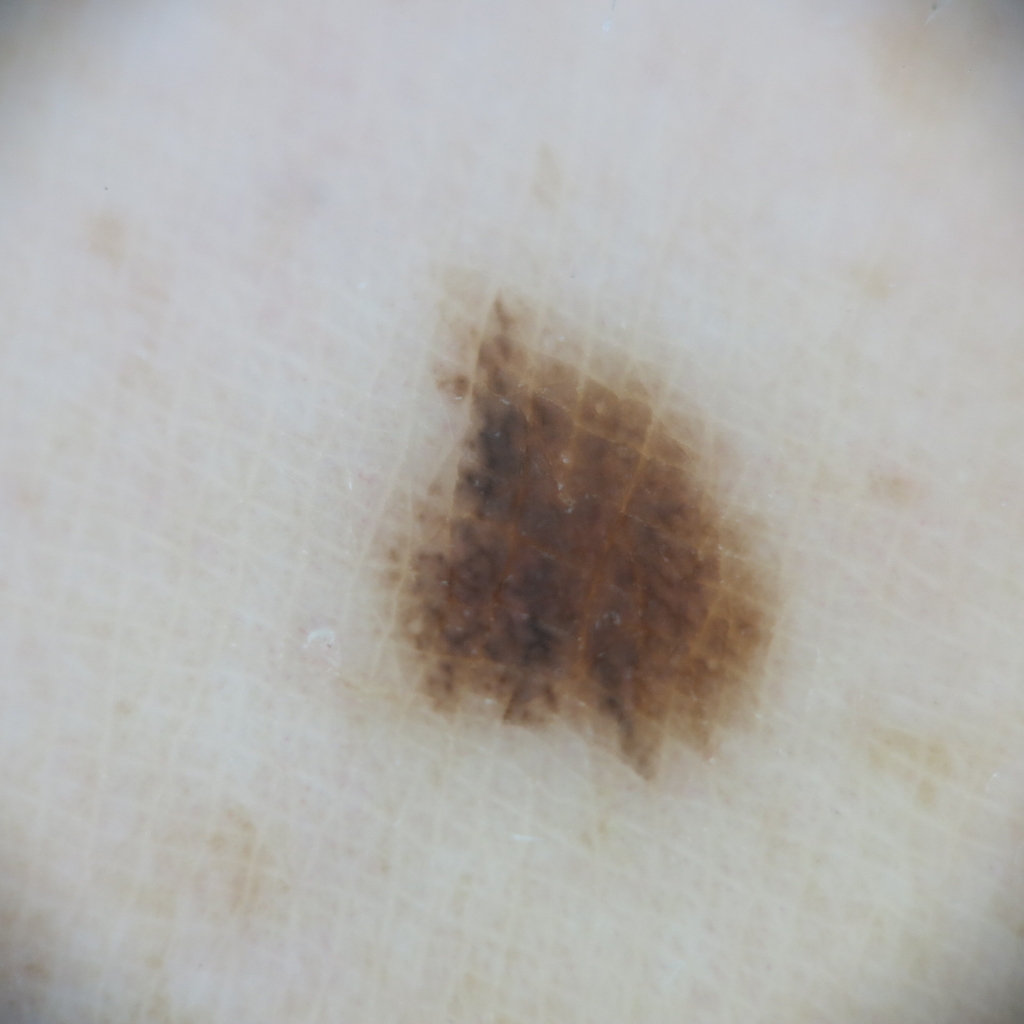}
        \centering
        (a)
      \endminipage
      \hfill
      \minipage{0.24\textwidth}    
        \includegraphics[width=\linewidth]{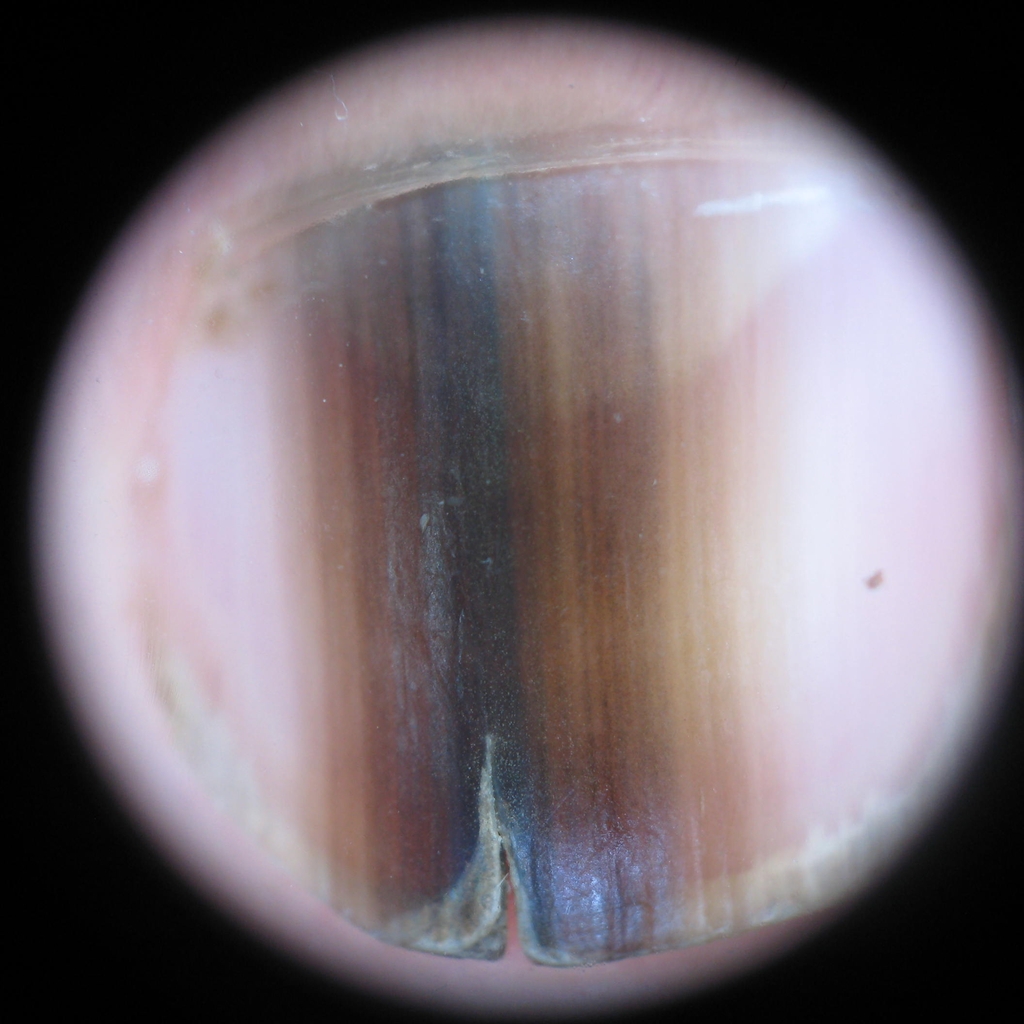}
        \centering
        (b)
      \endminipage
      \hfill
      \minipage{0.24\textwidth}    
        \includegraphics[width=\linewidth]{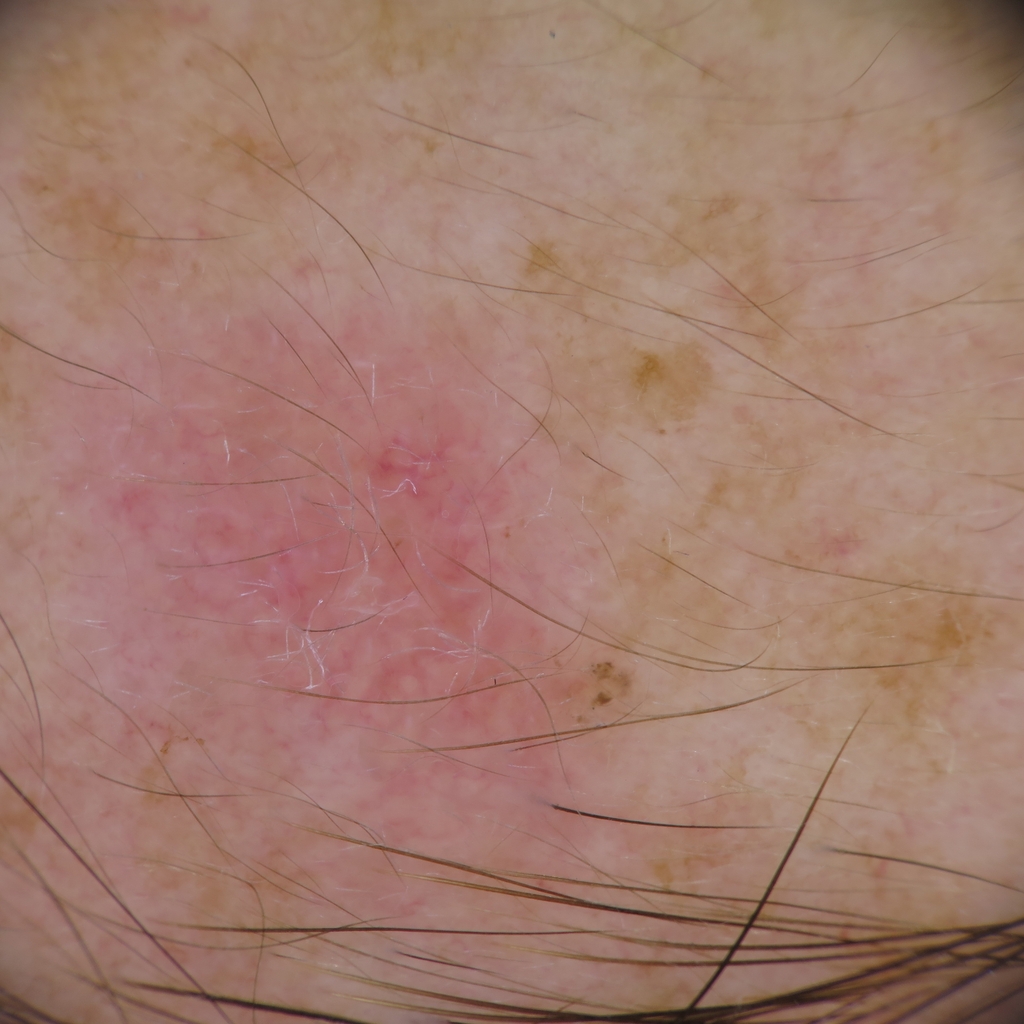}
        \centering
        (c)
      \endminipage
      \hfill
      \minipage{0.24\textwidth}    
        \includegraphics[width=\linewidth]{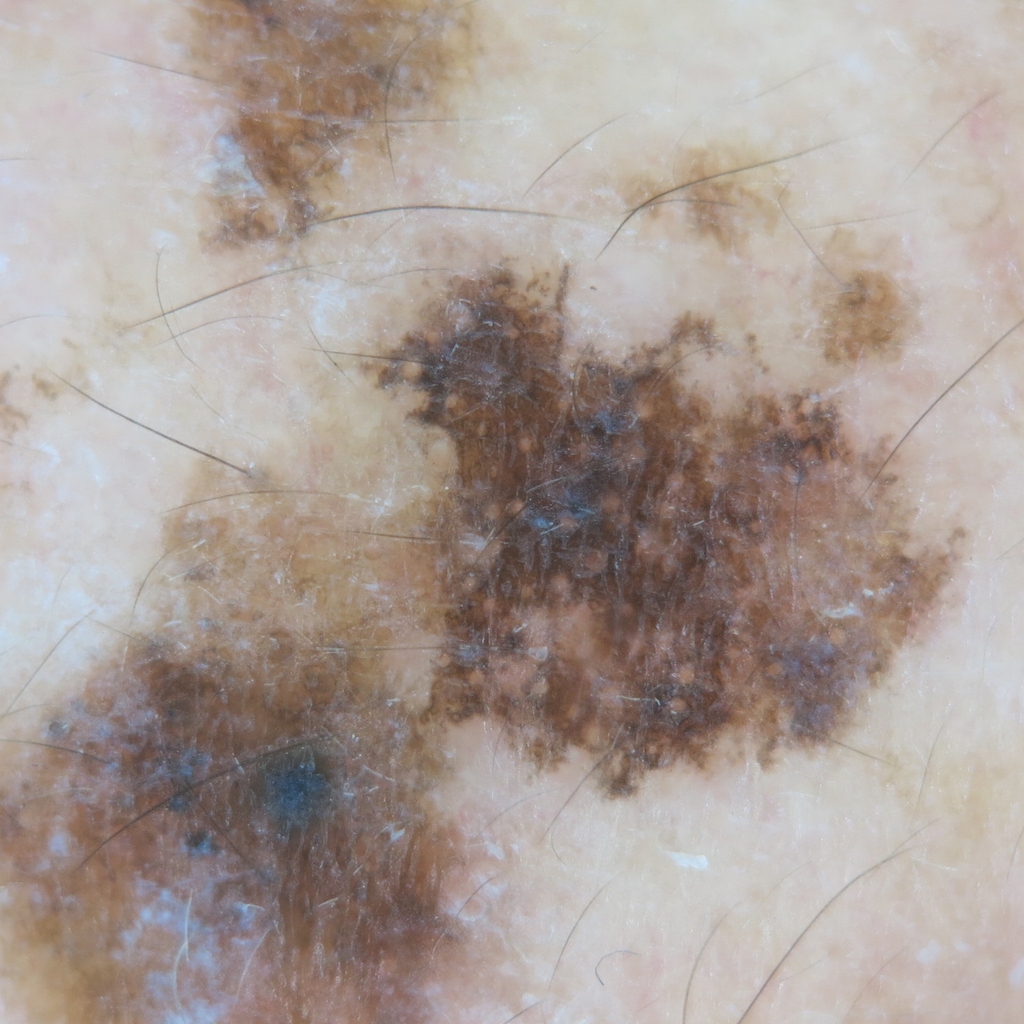}
        \centering
        (d)
      \endminipage
    \endminipage
    \vspace{0.1cm}
    \minipage{1\textwidth}
      \minipage{0.24\textwidth}
        \includegraphics[width=\linewidth]{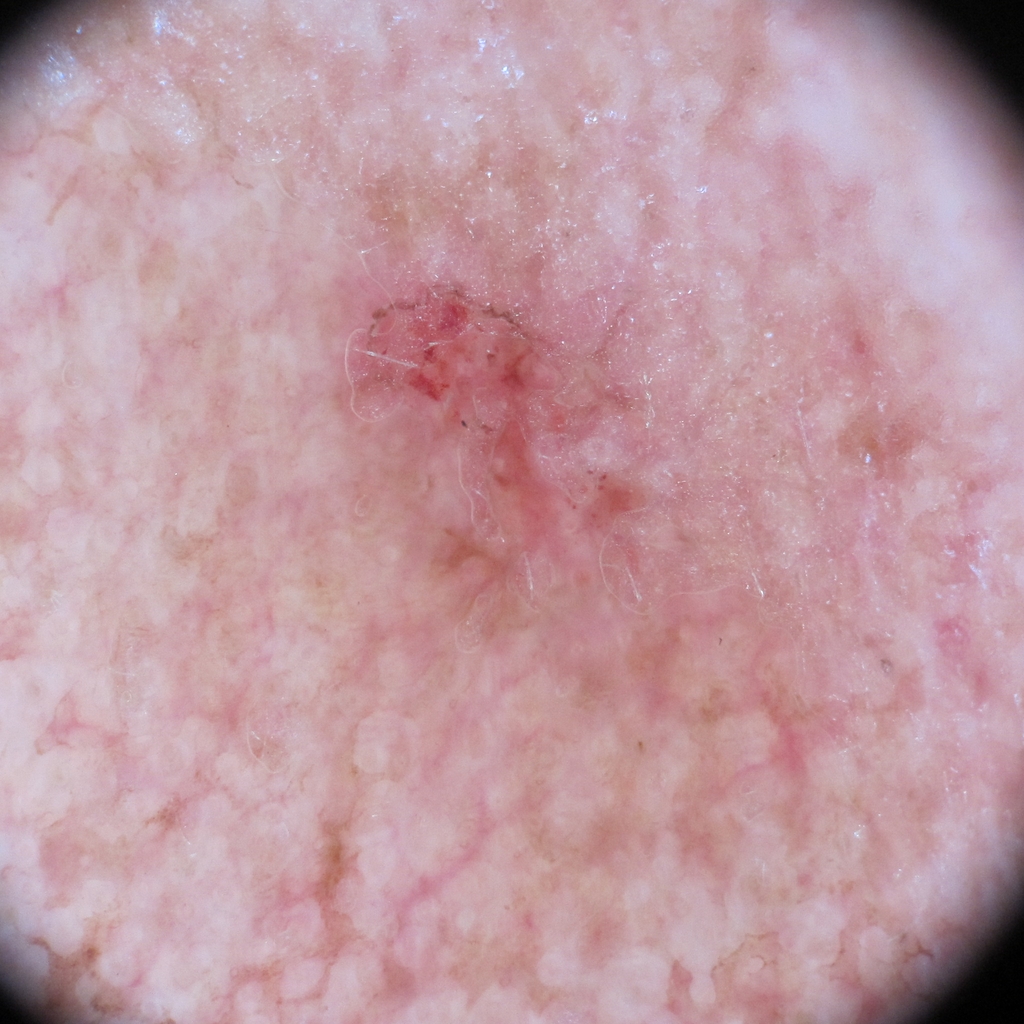}
        \centering
        (e)
      \endminipage
      \hfill
      \minipage{0.24\textwidth}    
        \includegraphics[width=\linewidth]{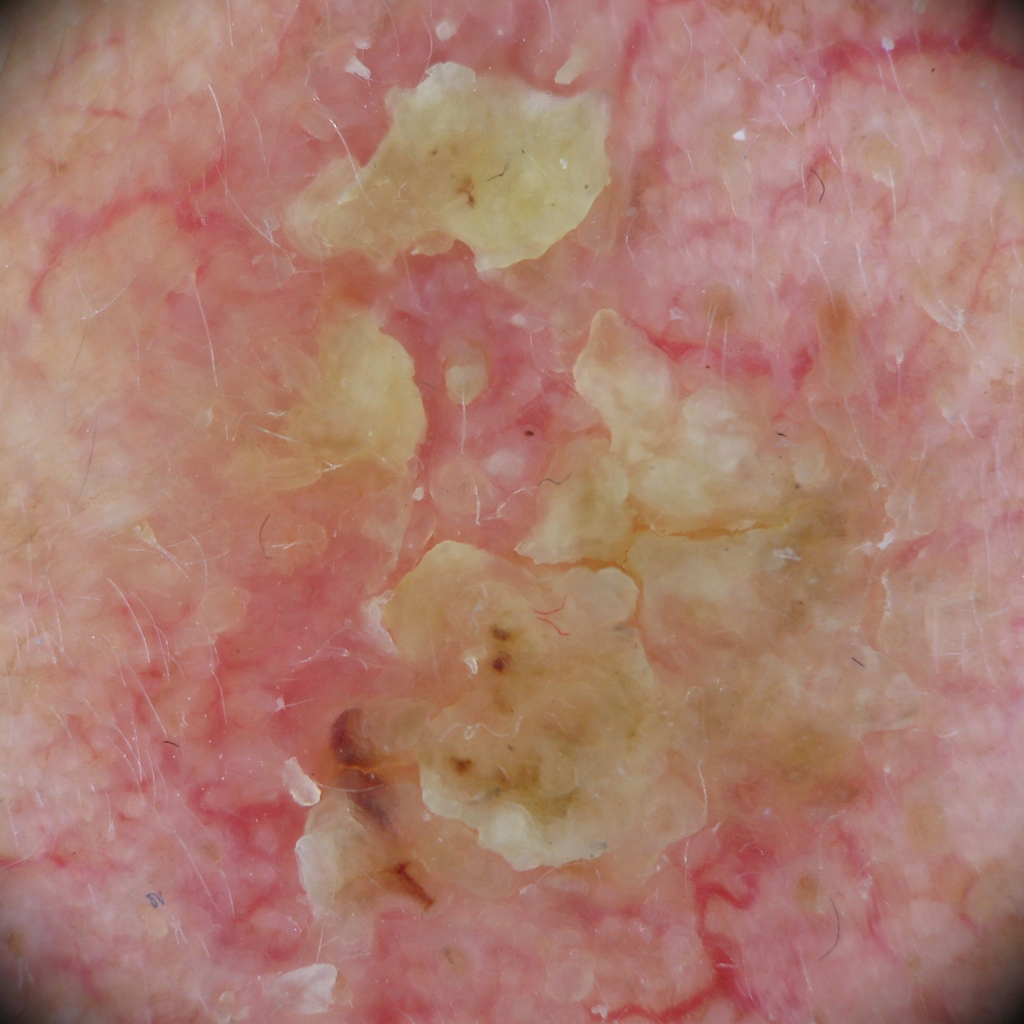}
        \centering
        (f)
      \endminipage
      \hfill
      \minipage{0.24\textwidth}    
        \includegraphics[width=\linewidth]{figures/dermatofibroma.jpg}
        \centering
        (g)
      \endminipage
      \hfill
      \minipage{0.24\textwidth}    
        \includegraphics[width=\linewidth]{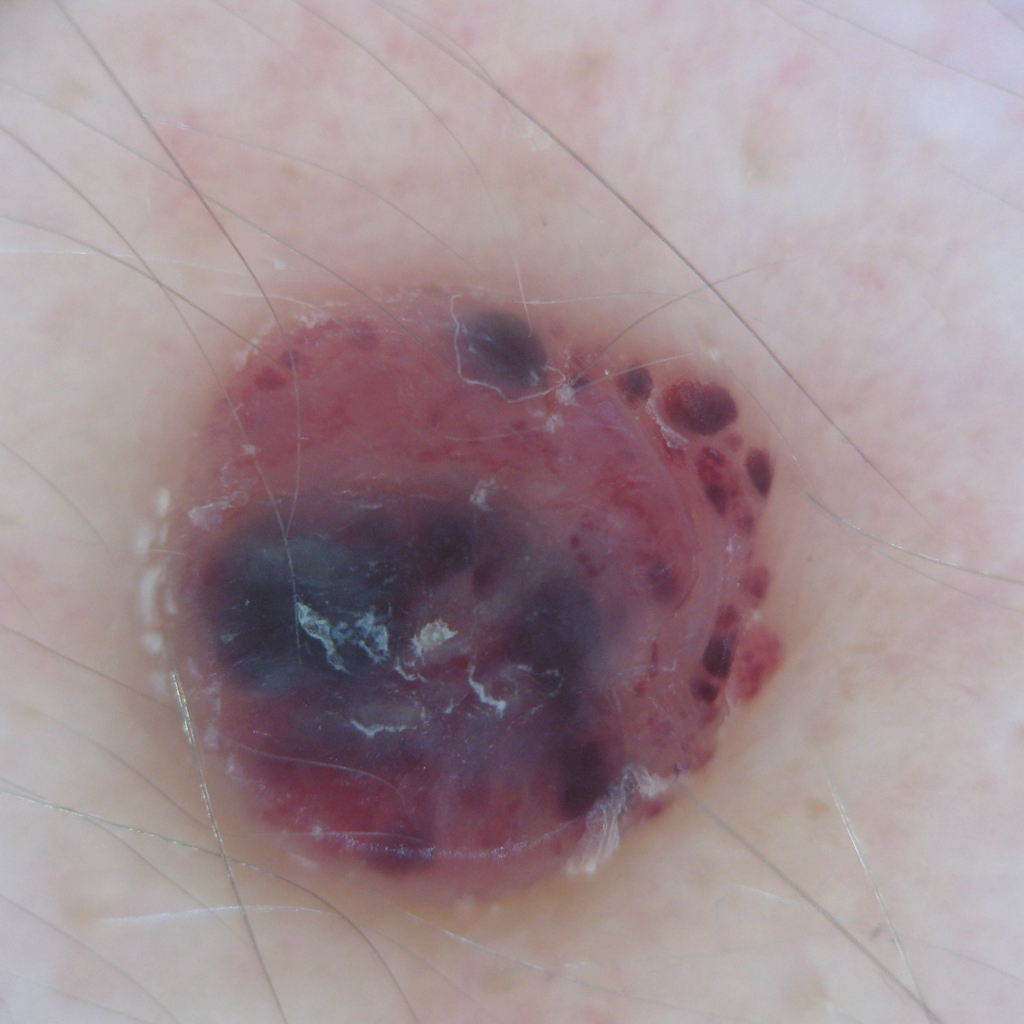}
        \centering
        (h)
      \endminipage
    \endminipage
\endminipage
\caption{Samples from the BCN20000 dataset correspodning to (a) nevus, (b) melanoma, (c) basal cell carcinoma, (d) seborrheic keratosis, (e) actinic keratosis, (f) squamos cell carcinoma, (g) dermatofibroma and (h) vascular lesion.}
\end{figure}

\section{Methods}

During more than 16 years, the Department of Dermatology at the “Hospital Clínic de Barcelona” has been systematically collecting dermoscopic images of skin lesions on their patients. The BCN20000 includes the dermoscopic images captured from 2010 until 2016 using a set of dermoscopic attachments on three high-resolution cameras that were stored using a directory structure in a server of the hospital. In order to create the BCN20000 database, these images have been retrieved, organized and filtered using various computer vision algorithms. Then, they have been linked with their corresponding diagnoses using a reference database. Finally, they have been manually revised to reassure plausibility of the diagnosis by several readers. The resulting database includes 19424 dermoscopic high-quality images corresponding to 5583 skin lesions captured between 2010 and 2016.
All the data contained in the BCN20000 database has received the necessary institutional ethics approval (HCB/2019/0413).

\begin{figure}[!htb]
    \centering
    \includegraphics[width=0.6\linewidth]{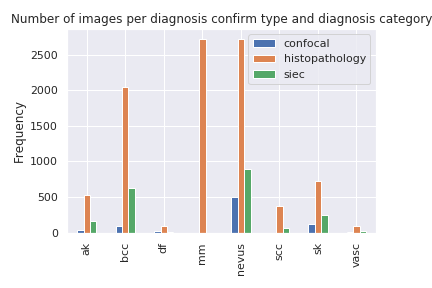}
    \caption{Image count for each diagnosis confirm type (siec: single image expert consensus).}
\end{figure}

\section{Usage Notes}

The images from the BCN20000 database can be divided into the following categories: nevus, melanoma, basal cell carcinoma, seborrheic keratosis, actinic keratosis, squamos cell carcinoma, dermatofibroma, vascular lesion and 'other' (lesions not contained in any of the other categories). To make the task more similar to clinical routine, each image is coupled with metadata regarding the anatomic location of the lesion, and the age and sex of the patient.

The dataset will be part of the ISIC 2019 Challenge \cite{ISIC2019}, where participants will be asked to classify among various diagnostic categories and identify out of the distribution situations, where the algorithm is seeing a skin lesion it has not been trained to deal with. We will also make the dataset available through the ISIC Archive \cite{ISICArchive}. 

\small
\bibliographystyle{abbrv}
\bibliography{mybib}{}

\end{document}